\documentclass[final, aps, prb, citeautoscript, twocolumn, superscriptaddress,floatfix]{revtex4-1}

\usepackage{graphicx} 
\usepackage{amsmath}
\usepackage{amsfonts}
\usepackage{amssymb}
\usepackage{nccmath} 

\newcommand{\rY}[2]{R_{#1}^{#2}}
\newcommand{\zY}[2]{Y_{#1}^{#2}}
\newcommand{\sY}[3]{S_{{#1}{#2}{#3}}}

\newcommand{\rpair}{{\,p}}

\newcommand{\xn}{\bar{x}}
\newcommand{\yn}{\bar{y}}
\newcommand{\zn}{\bar{z}}

\usepackage{bm}

\allowdisplaybreaks

\usepackage{acro}
\acsetup{barriers/use=true, barriers/reset=true}

\DeclareAcronym{CPU}{
short = {CPU},
long = {central processing unit}
}

\DeclareAcronym{DFT}{
short = {DFT},
long = {density functional theory}
}

\DeclareAcronym{GPL}{
short = {GPL},
long = {GNU public license}
}

\DeclareAcronym{GPU}{
short = {GPU},
long = {graphics processing unit}
}

\DeclareAcronym{GSL}{
short = {GSL},
long = {the GNU scientific library}
}
\DeclareAcronym{PAW}{
short = {PAW},
long = {projector-augmented wave}
}

\DeclareAcronym{stdc++}{
short = {stdc++},
long = {the standard C++ library}
}

\usepackage{pgfplots}
\pgfplotsset{compat=1.9}
\usepackage[dvipsnames,svgnames]{xcolor}
\usetikzlibrary{patterns}

\newcommand{\llnl}{Quantum Simulations Group, Lawrence Livermore National Laboratory, Livermore, California 94550, USA}
\newcommand{\madison}{Department of Materials Science and Engineering, University of Wisconsin, Madison, WI, 53706, USA}

\begin{document}

\title{SHarmonic: A fast and accurate implementation of spherical harmonics for electronic-structure calculations}

\author{Xavier Andrade}
\email{xavier@llnl.gov}
\affiliation{\llnl}

\author{Jacopo Simoni}
\affiliation{\madison}

\author{Yuan Ping}
\affiliation{\madison}
\affiliation{Department of Physics, University of Wisconsin, Madison, WI, 53706, USA}
\affiliation{Department of Chemistry, University of Wisconsin, Madison, WI, 53706, USA}

\author{Tadashi Ogitsu}
\affiliation{\llnl}

\author{Alfredo A. Correa}
\affiliation{\llnl}

\begin{abstract}
The authors present SHarmonic, a new implementation of the spherical harmonics targeted for electronic-structure calculations.
Their approach is to use explicit formulas for the harmonics written in terms of normalized Cartesian coordinates.
This approach results in a code that is as precise as other implementations while being at least one order of magnitude more computationally efficient.
The library can run on \acp{GPU} as well, achieving an additional order of magnitude in execution speed.
This new implementation is simple to use and is provided under an open source license, it can be readily used by other codes to avoid the error-prone and cumbersome implementation of the spherical harmonics.
\end{abstract}

\maketitle 

\acbarrier

\section{Introduction}

The spherical harmonics are a set of special functions that are crucial for understanding the atoms at the quantum mechanical level.
As such, the numerical calculation of these functions is required in most electronic structure codes, for example for atomic-orbital basis sets or the projectors in non-local pseudo-potentials.

In this article we present SHarmonic, an accurate and fast implementation of the spherical harmonics.
The code was originally developed for the INQ code~\cite{Andrade2021} and it is now a standalone header only library that can be used from C and C++.
SHarmonic approach is to directly implement the spherical harmonics formulas up to order 9.
We show this approach produces very good results in terms of precision and numerical performance.

We present and release this library in the interest of code reusability~\cite{Oliveira2020}, and reproducibility~\cite{Lejaeghere2016}.
We also provide the set of explicit formulas for the spherical harmonics as implemented in our code, and reference values to validate spherical harmonic codes.

In pseudopotential~\cite{Hellmann1935, Hamann1979, Vanderbilt1990, Hamann2013} and \ac{PAW}~\cite{Blochl1994,Kresse1999} implementations of \ac{DFT} the atomic problem is solved in a radial grid for each species or each atom.
This produces a series of angular-momentum dependent projectors~\cite{Kleinman1982}.
To represents these projectors in a 3-dimensional grid we need to calculate the spherical harmonics for each point.
It is in this step that the spherical harmonics are required in plane-wave or real-space electronic structure codes.

Codes that rely on atomic-orbital basis-sets~\cite{Ditchfield1971, Soler2002, VanLenthe2003, Blum2009, Jensen2012} may also use the spherical harmonics as the radial part of the basis set. Even though in some cases a simpler, but larger, Cartesian polynomial basis of the form \(x^{\alpha}y^{\beta}z^{\gamma}\) is used~\cite{Schlegel1995}.

The calculation of the spherical harmonics is usually not a computational bottleneck.
However, it is important to have an implementation that is fast enough compared to the rest of the code to avoid the computational cost to become significant.
This is particularly important when running on \acp{GPU}\cite{Andrade2013}. If the spherical harmonics are still calculated on the GPU, their fraction of the computational time will increase significantly, and more importantly, this might require copying data between the \ac{CPU} and \ac{GPU}.

\section{The spherical harmonics}

The spherical harmonics \(\zY{\ell}{m}\) are a set of complex functions defined on the surface of a sphere~\cite{Arfken2013angular}.
In quantum mechanics they appear as the eigen-functions of the angular momentum operators \(\hat{L}^2\) and \(L_z\).
The two indices \(\ell\) and \(m\) are related to the eigenvalues of these operators (atomic units are used throughout)
\begin{align}
    \hat{L}^2\,\zY{\ell}{m} &= (\ell + 1)\ell\,\zY{\ell}{m}\\
    \hat{L}_z\,\zY{\ell}{m} & = m\,\zY{\ell}{m} 
\end{align}

As functions on a sphere, the spherical harmonics are naturally defined in terms of the angular part of the spherical coordinates, given by the polar angle \(\theta\) and azimuthal angle \(\phi\). In terms of standard Cartesian coordinates \(x, y, z\), these angles are defined as
\begin{align}
\theta & = \mathrm{acos}\left(\frac{z}{r}\right)\ ,\label{eq:spherical_theta}\\
\phi   & = \mathrm{atan2}(y,x)\ .
\label{eq:spherical_phi}
\end{align}
Note that some conventions invert the definitions of \(\theta\) and \(\phi\).
In these angular coordinates, the explicit form of the spherical harmonics is
\begin{equation}
    \zY{\ell}{m} = (-1)^m\, 
        \sqrt{\frac{2\ell + 1}{4\pi}\frac{(\ell - m)!}{(\ell + m)!}}
        P^\ell_m(\cos\theta)\,e^{im\phi}
        \label{eq:sphericalfromlegendre}
\end{equation}
where \(P^\ell_m\) are the associated Legendre polynomials~\cite{Arfken2013legendre}.
We include Condon–Shortley phase, \((-1)^m\), in the definition and implementation of the spherical harmonics~\cite{Condon1935}.

For non-relativistic electronic structure calculations the atomic wave-functions can be chose to be real valued. It is convenient then to use the real-valued form of the spherical harmonics, defined as
\begin{align}
    \rY{\ell}{m} &= \begin{cases}
\dfrac{i}{\sqrt{2}} \left(\zY{\ell}{m} - (-1)^m\, \zY{\ell}{-m}\right) & \text{if}\ m < 0\\
Y_\ell^0 & \text{if}\ m=0\\
\dfrac{1}{\sqrt{2}} \left(\zY{\ell}{-m} + (-1)^m\, \zY{\ell}{m}\right) & \text{if}\ m > 0\ .
\end{cases}
\label{eq:real_harmomics}
\end{align}

For relativistic electronic-structure calculations~\cite{Eschrig2004,Truhlar2025,Simoni2025} it is necessary to use the \emph{spinorial} version of spherical harmonics.
These are two-component spinors that depend on the total angular momentum: orbital plus spin. 
These values are represented by the quantum numbers \(j\) and \(m_j\).
For spin \(\frac{1}{2}\) the \emph{spinor} spherical harmonics are defined as
\begin{multline}
\label{eq:spinorial}
  \sY{j}{m_j}{\pm\frac{1}{2}}
    =\\ \frac{1}{\sqrt{2 \bigl(j \mp \frac{1}{2}\bigr) + 1}}
      \begin{pmatrix}
        \pm \sqrt{j \mp \frac{1}{2} \pm m_j + \frac{1}{2}} \zY{j\mp\frac{1}{2}}{m_j - \frac{1}{2}} \\
        \sqrt{j \mp \frac{1}{2} \mp m_j + \frac{1}{2}} \zY{j\mp\frac{1}{2}}{m_j + \frac{1}{2}}
       \end{pmatrix}.
\end{multline}

For maximum flexibility, the SHarmonic library implements of all of these versions of the spherical harmonics: real, complex and spinorial.

In eq.~(\ref{eq:sphericalfromlegendre}) we define the spherical harmonics in term of spherical coordinates.
This is a logical choice since we can drop the dependence in \(r\) (with \(r = \sqrt{x^2 + y^2 + z^2}\)) and only consider the angular coordinates.
It is also natural when working on a single atom, that naturally defines a center for the spherical coordinates.

However, for electronic-structure codes, that usually simulate multiple atoms, it is convenient to define the spherical harmonics in term of Cartesian coordinates.
In this case the value of the functions is independent of the norm of the coordinates.
For \(r=0\), where the angular components are not well defined, we use the convention that the spherical harmonics are 0, except for \(l = 0\) that is a constant function.

We can use the independence on the norm to define the spherical coordinates in terms of \emph{normalized} Cartesian coordinates.
These coordinates are represented by three values, \(\xn\), \(\yn\) and \(\zn\) with the constraint that
\begin{equation}
\label{eq:constraint}
\xn^2 + \yn^2 + \zn^2 = 1\ .
\end{equation}
In these coordinates the spherical harmonics take a simple form as polynomial of the coordinate variables.
Also, we show that the normalization condition,~eq.~(\ref{eq:constraint})can be used to simplify the equations.
These facts make normalized Cartesian coordinates ideal for the numerical implementation of the spherical harmonics in explicit form, so this is the approach that SHarmonic takes.

For convenience for the users, SHarmonic provides interfaces in the three coordinate systems: angular, Cartesian and normalized Cartesian.
The later is the preferred interface as it offers the highest performance; it should be used if normalized coordinates or the value of \(r\) are available.
The other interfaces require the relatively-costly calculation of trigonometric functions, divisions and square roots.


\section{Implementation}

Despite its use in most electronic structure codes, there is no standard implementation of the spherical harmonics.
Many major electronic-structures codes use their own implementation.
There are also some libraries that provide the calculation of the spherical harmonics, or at least the associated Legendre polynomials: \ac{stdc++}, Boost, and \ac{GSL}.
We use three of these libraries as references for our implementation in terms of accuracy and performance.
We discuss these implementations in detail in Appendix~\ref{sec:alternatives}.
The main limitation of all of these alternatives is that they do not work on \acp{GPU}.

Our implementation of the spherical harmonics is done directly in normalized Cartesian coordinates using explicit formulas for each value of \(\ell\) and \(m\).
This results in a simple implementation as the spherical harmonics are simple polynomial functions that can be evaluated quickly on both the \ac{CPU} and \ac{GPU} using only additions and multiplications.
The main drawback of this explicit approach is that we have to limit our code up to a certain value of \(\ell\).
In this case we pick \(\ell=9\) as the higher order we implement, since our target are electronic-structure calculations this is a more than reasonable upper limit.

We also take several measures to restrict the number of formulas we need to implement.
To start, we only implement the formulas for the real spherical harmonics.
The \(m\ge0\) complex spherical harmonics can be calculated in terms of the real ones as
\begin{equation}
        \zY{\ell}{m} = (-1)^{m}\left[\rY{\ell}{m} + i\rY{\ell}{-m}\right]\ .
\end{equation}
While for negative values of \(m\) we use that
\begin{equation}
\label{eq:conjm}
        \zY{\ell}{-m} = (-1)^{m}(\zY{\ell}{m})^*\ .
\end{equation}
The spinorial harmonics are calculated in terms of the complex harmonics using~eq.~(\ref{eq:spinorial}).

To further reduce the size of our implementation, by almost a factor of two, we only explicitly implement the formulas for the \(m\le0\) real spherical harmonics.
While eq.~(\ref{eq:conjm}) does not hold for real harmonics, we have a derived formulas to connect \(\rY{\ell}{m}\) with \(\rY{\ell}{-m}\) through coordinate rotations.
The details are in Appendix~\ref{sec:mconnection}.

To obtain the explicit formulas for the spherical harmonics we use a third-party script based on computational symbolic algebra to calculate the explicit expressions for the real spherical harmonic in Cartesian coordinates~\cite{Maeda2025}.
For example, for the \(\ell=7\) and \(m=-3\) spherical harmonic we obtain
\begin{multline}
\label{eq:r37a}
 \rY{7}{-3} =
 128\sqrt{\frac{35}{\pi}}\yn(2574\,\xn^6 + 4290\xn^4\yn^2 + 858\,\xn^2\yn^4 - 858\,\yn^6\\
  - 3960\,\xn^4 - 2640\,\xn^2\yn^2 + 1320\,\yn^4  + 1440\,\xn^2 - 480\,\yn^2)\ .
\end{multline}

The polynomials we obtain from the script are written in canonical form. This form is slow to evaluate and can result in large numerical errors in finite-precision calculations. 
This makes it necessary to rewrite them in a form that is more amenable to numerical calculations.

The standard approach for the evaluation of single-variable polynomials is to use the Horner's form~\cite{Horner1819}.
However, this is not useful for multi-variate polynomials.
We use a different approach by writing the polynomials in fully factorized form in terms of their roots.
To achieve this form, for each polynomial we find the most convenient combination of different transformations. These transformations are done using the online symbolic algebra package Wolfram Alpha~\cite{WolframAlpha2024}.

To demonstrate this procedure, we show the transformations we apply to the expression for \(\rY{7}{-3}\) in eq.~(\ref{eq:r37a}).
Note, however, that for each polynomial, the actual combination of transformations can be different.
The most basic operation is to directly factorize the polynomial, which gives us
\begin{multline}
\label{eq:r37b}
 \rY{7}{-3} =
 768\sqrt{\frac{35}{\pi}}\yn(3\,\xn^2 - \yn^2)(143\,\xn^4 + 286\,\xn^2 \yn^2\\ 
 + 143\,\yn^4 - 220\,\xn^2 - 220\,\yn^2 + 80)\ .
\end{multline}
In this case, the second term cannot be factorized further in a simple form.
However, we can factorize the polynomial ignoring the constant term, to obtain
\begin{multline}
\label{eq:r37c}
 \rY{7}{-3} =
 768\sqrt{\frac{35}{\pi}}\yn(3\,\xn^2 - \yn^2)\\
 \times(11\,(\xn^2 + \yn^2) (13\,\xn^2 + 13\,\yn^2 - 20) + 80)\ .
\end{multline}
We can now use the normalization condition, eq~(\ref{eq:constraint}),  that implies that \(\xn^2 + \yn^2 = 1 - \zn^2\).
This allows us to convert multi-variate polynomials in \(\xn\) and \(\yn\) into single-variate polynomials in \(\zn\).
We have found this is always possible to do with non-homogeneous polynomials in \(\xn\) and \(\yn\) that appear in the spherical harmonics.
This is not surprising since \(\xn^2 + \yn^2\) in angular coordinates is \(\sin^2\theta\).
This procedure applied to eq.~(\ref{eq:r37c}) yields.
\begin{multline}
 \rY{7}{-3} =
 768\sqrt{\frac{35}{\pi}}\yn(3\,\xn^2 - \yn^2)(11 (\zn^2 - 1) (13\,\zn^2 + 7) + 80)\ .
\end{multline}
This new polynomial in \(\bar{z}\) can be written in Horner's form, or in this case we find its analytical roots.
The final expression for this particular spherical harmonic is
\begin{multline}
  \rY{7}{-3}	  = \frac{429}{64}\sqrt{\frac{35}{\pi}}\yn\rpair\left(\sqrt{3}\,\xn, \yn\right)\\
  \rpair\left(\sqrt{\frac{33 - 2\sqrt{165}}{143}}, \zn\right)\rpair\left(\sqrt{\frac{33 + 2\sqrt{165}}{143}}, \zn\right)\ ,
 \end{multline}
where, to simplify the notation and implementation, we define the function \(\rpair\) as
\begin{equation}
  \label{eq:rpair}
  \rpair(a, b) = (a - b)(a + b)\ .
\end{equation}
This function represents the polynomial factors associated with a pair of positive and negative roots with the same absolute value. Note that while it might be tempting to simply write \(\rpair(a,b)\) as \(a^2-b^2\), this can result in larger round-off errors when \(|a|\) is close to \(|b|\).

We repeat this procedure for each spherical harmonic up to \(\ell = 9\) trying to find the best strategy to get the best expression in terms of numerical accuracy and performance.
In some cases, the roots do not have a simple analytical expression, so we either write the polynomial in Horner's form or we use high precision numerical values of the roots.
The resulting formulas for all the harmonics are given in Appendix~\ref{sec:formulas}.


\section{Accuracy}

For practical uses, it is essential that our implementation of the spherical harmonics is as accurate as the alternatives.
To test it we compare the result of our code with three established libraries that can be used to implement the spherical harmonics: \ac{stdc++} , Boost, and \ac{GSL}.
We discuss these reference libraries in detail in Appendix~\ref{sec:alternatives}.

For this comparison we generate 5180 random points distributed uniformly inside a sphere.
We calculate \(\theta\) and \(\phi\) for each point and we pass these values to our implementation and to the reference implementations. We compare the resulting values of the real spherical harmonics.
We use angular coordinates since the reference libraries are implemented in terms of them, so we want to avoid small differences that might appear in the conversion, especially close to the \(z\) axis where the \(\phi\) angle is not well defined.

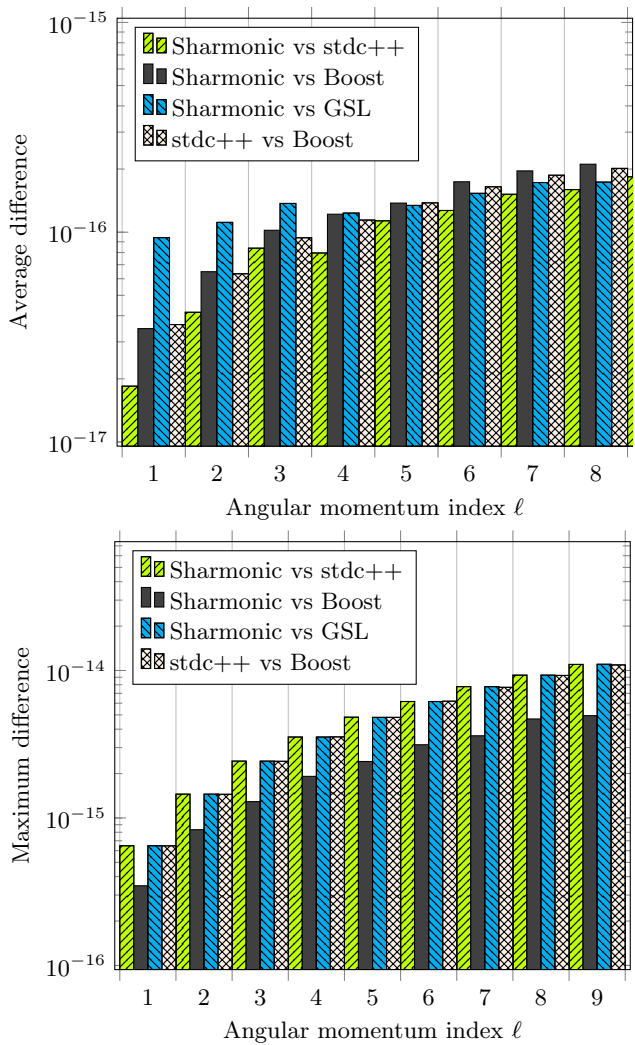
\begin{figure}
    \centering
     \begin{tikzpicture}
   \begin{axis}[
    ybar interval=1,
    enlargelimits=0.01,
    legend style={at={(0.31,0.98)}, anchor=north},
    legend cell align=left,
    ymode=log,
    log origin=infty,
    ymin=1e-17,
    ymax=1e-15,
    ylabel={Average difference},
    xlabel={Angular momentum index \(\ell\)},
     ]
    \pgfplotstableread{
l sh_v_std sh_v_boo sh_v_gsl std_v_boo std_v_gsl
1	1.84533e-17	3.47018e-17	9.42207e-17	3.62962e-17	8.19521e-17	
2	4.15136e-17	6.48147e-17	1.11393e-16	6.33551e-17	9.19894e-17	
3	8.39157e-17	1.02139e-16	1.37108e-16	9.40664e-17	8.38775e-17	
4	7.96196e-17	1.21855e-16	1.23367e-16	1.14265e-16	7.72825e-17	
5	1.13279e-16	1.37668e-16	1.34164e-16	1.38018e-16	7.43413e-17	
6	1.26913e-16	1.73916e-16	1.53207e-16	1.64652e-16	7.68652e-17	
7	1.5162e-16	1.96027e-16	1.72529e-16	1.86895e-16	7.93798e-17	
8	1.59642e-16	2.10921e-16	1.73317e-16	2.01443e-16	8.59813e-17	
9	1.83695e-16	2.44714e-16	1.95088e-16	2.24174e-16	9.06702e-17	
10	1.09912e-14	4.94049e-15	1.1019e-14	1.09079e-14	5.32907e-15
    }\loadedtable
    \addplot[black,fill=lime,postaction={
      pattern=north east lines
    }] table [x=l,y=sh_v_std] {\loadedtable};
    \addplot[black,fill=darkgray] table [x=l,y=sh_v_boo] {\loadedtable};
    \addplot[black,fill=cyan,postaction={
      pattern=north west lines
    }] table [x=l,y=sh_v_gsl] {\loadedtable};
    \addplot[black,fill=Linen,postaction={
      pattern=crosshatch
    }] table [x=l,y=std_v_boo] {\loadedtable};    
    \legend{
      {Sharmonic vs stdc++},
      {Sharmonic vs Boost},
      {Sharmonic vs GSL},
      {stdc++ vs Boost}}
\end{axis}
\end{tikzpicture}%
\\
 \begin{tikzpicture}
   \begin{axis}[
     ybar interval=1,
     enlargelimits=0.01,
     legend style={at={(0.31,0.98)}, anchor=north},
     legend cell align=left,
     ymode=log,
     log origin=infty,
     ymin=1e-16,
     ymax=7e-14,
     ylabel={Maximum difference},
     xlabel={Angular momentum index \(\ell\)},
     ]
    \pgfplotstableread{
l sh_v_std sh_v_boo sh_v_gsl std_v_boo std_v_gsl
1	6.47919e-16	3.46945e-16	6.47919e-16	6.47052e-16	3.33067e-16	
2	1.45023e-15	8.32667e-16	1.45023e-15	1.44676e-15	5.55112e-16	
3	2.42861e-15	1.29063e-15	2.42861e-15	2.42167e-15	6.66134e-16	
4	3.53884e-15	1.91513e-15	3.53884e-15	3.54577e-15	1.11022e-15	
5	4.82253e-15	2.41474e-15	4.80865e-15	4.80865e-15	1.88738e-15	
6	6.16174e-15	3.13638e-15	6.16174e-15	6.18949e-15	2.66454e-15	
7	7.77156e-15	3.60822e-15	7.77156e-15	7.68829e-15	3.66374e-15	
8	9.29812e-15	4.69069e-15	9.29812e-15	9.25648e-15	4.55191e-15	
9	1.09912e-14	4.94049e-15	1.1019e-14	1.09079e-14	5.32907e-15	
10	1.09912e-14	4.94049e-15	1.1019e-14	1.09079e-14	5.32907e-15
    }\loadedtable    
    \addplot[black,fill=lime,postaction={
      pattern=north east lines
    }] table [x=l,y=sh_v_std] {\loadedtable};
    \addplot[black,fill=darkgray] table [x=l,y=sh_v_boo] {\loadedtable};
    \addplot[black,fill=cyan,postaction={
      pattern=north west lines
    }] table [x=l,y=sh_v_gsl] {\loadedtable};
    \addplot[black,fill=Linen,postaction={
      pattern=crosshatch
    }] table [x=l,y=std_v_boo] {\loadedtable};    
    \legend{
      {Sharmonic vs stdc++},
      {Sharmonic vs Boost},
      {Sharmonic vs GSL},
      {stdc++ vs Boost}}
\end{axis}
\end{tikzpicture}
    \caption{Comparison of the accuracy between SHarmonic and reference implementation of the real spherical harmonics.
    The comparison is done over a set of 5180 random values uniformly distributed over a sphere.
    Top panel: Difference between implementation for each \(\ell\) averaged over points and \(m\).
    Bottom panel: Maximum difference between implementations for each \(\ell\). 
    This results show that SHarmonic is as accurate as the reference implementations.
    }
    \label{fig:accuracy}
\end{figure}

Our comparison results are shown in fig.~\ref{fig:accuracy}.
We find that on average SHarmonic is very close to the reference implementations, with average differences on the order of the machine precision, \(10^{-16}\).
When looking at the maximum difference among all the points we tested, the difference goes up to \(10^{-14}\).
We found that the larger difference happens on points with values where \(\theta\) is close to zero.
Similar differences happen between reference implementations as well, with \ac{stdc++} and Boost showing similar maximum differences of \(10^{-14}\).
Based on this comparison, we can conclude our code is as accurate in the calculation of the spherical harmonics as the other implementations we tested.


\section{Numerical performance}

\begin{table}[ht]
    \centering
    \begin{tabular}{lrrrr}
        \hline
        & \multicolumn{2}{c}{CPU} & \multicolumn{2}{c}{GPU} \\
        \cline{2-3} \cline{4-5}
         Input coordinates & Real & Complex & Real & Complex \\
        \hline
        Cartesian & 228 & 85 & 9,939 & 5,187 \\
        Normalized Cartesian & 553 & 154 & 12,540 & 5,143 \\
        Angular & 159 & 70 & 9,307 & 5,305 \\
        \hline
    \end{tabular}
    \caption{Throughput in millions of harmonics per seconds for different SHarmonic functions running on CPU and GPU. Execution on a single core of a AMD Ryzen 5950x \ac{CPU} and an Nvidia V100 \ac{GPU}.}
    \label{tab:cpu_gpu_comparison}
\end{table}

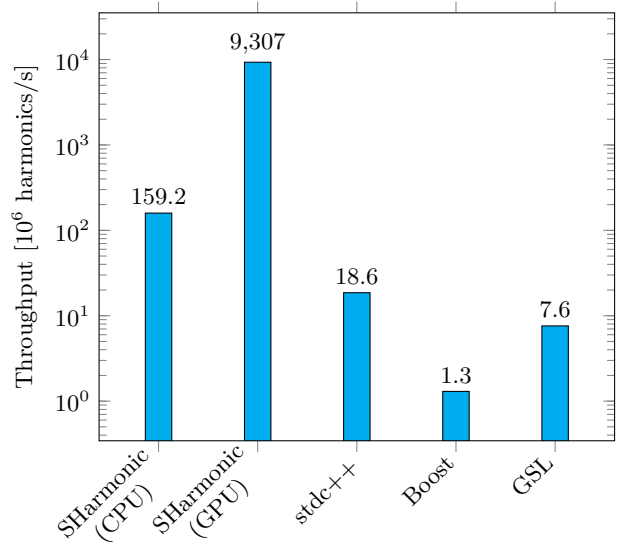
\begin{figure}[ht]
    \centering
\begin{tikzpicture}
\begin{axis}[
    ybar,
    enlargelimits=0.15,
    legend style={at={(0.5,-0.15)},
      anchor=north,legend columns=-1},
    ylabel={Throughput [\(10^6\) harmonics/s]},
    symbolic x coords={SHarmonic (CPU), SHarmonic (GPU), stdc++, Boost, GSL},
    xticklabels={\\SHarmonic\\ (CPU), SHarmonic\\ (GPU), \\stdc++, Boost, GSL},
    xticklabel style={align=center},
    xtick=data,
    nodes near coords,
    nodes near coords align={vertical},
    point meta=rawy,
    ymode=log,
    log origin=infty,
     x tick label style={rotate=45,anchor=east}
    ]
\addplot[black,fill=cyan] coordinates {(SHarmonic (CPU), 159.2) (SHarmonic (GPU), 9307) (stdc++,18.6) (Boost, 1.3) (GSL, 7.6)};
\end{axis}
\end{tikzpicture}
    \caption{Comparson of the computational throughput in millions of harmonics per second for different implementations of the spherical harmonics: SHarmonic on CPU and GPU, . Calculation of real harmonics in angular coordinates averaged over \(\ell\), between 1 and 9, and \(m\). Execution on a single core of a AMD Ryzen 5950x \ac{CPU} and an Nvidia V100 \ac{GPU}.}
    \label{fig:performance}
\end{figure}

Now that we established our implementation is accurate we focus on the numerical performance.
To measure performance we calculate the spherical harmonics for a 100 points with a 1000 repetitions each (1,024,000 for GPU calculations).
To ensure the calculations are not optimized out by the compiler, we do a small variation in the input for each repetition and accumulate over the results.

We start by evaluating the numerical performance of our implementation for different processor types (CPU or GPU),  input coordinates, and harmonic type (real or complex).
The results are shown in table~\ref{tab:cpu_gpu_comparison}.
We can see that the GPU implementation can be up to two orders of magnitude faster than the CPU version.
Also, as expected the real harmonics are faster to compute than the complex ones, as the complex harmonics are implemented using two real harmonics.
While we expect a factor of 2 difference, we see that in some cases the real harmonics can be more than 3 times faster.
In terms of the input coordinates we find that for most cases the normalized Cartesian version of the functions is faster.
This is expected since this version does not need trigonometric functions or normalization.
However in the complex GPU case we do not see a variation with the input coordinate type.

We now compare the numerical performance of our implementation with respect to the reference libraries.
As shown in fig.~\ref{fig:performance}, when running on the same CPU SHarmonic is approximately two or three order of magnitude faster than the other implementations.
The fastest of the alternatives is the implementation based on the C++ standard library, and boost is the slowest.
When SHarmonic is running on the GPU we find an impressive speed of \(500\times\) with respect to the fastest reference library running on the CPU.


\section{Software distribution}

The purpose of writing and releasing SHarmonic is to provide the electronic-structure community with a readily available implementation of the spherical harmonics.
With this objective in mind we have designed SHarmonic to be easily integrated into existing and new codes, both from the legal and technical point of view.

SHarmonic is released under the Mozilla Public License 2.0, this is an open source license that guarantees every one access to the code and modified versions.
At the same time, this license guarantees that SHarmonic can be integrated into practically any open source or commercial code without licensing issues.

The library has interfaces for the C and C++ programming languages.
The C++ version has a more advanced interface that allows the user to decide what type of complex objects to use.
For the moment SHarmonic can be called from a Fortran code using the C interface, we expect to develop a native Fortran interface in the future.

The source code for SHarmonic can be found on \url{https://gitlab.com/npneq/sharmonic}.
To simplify distribution the library is designed to be header only.
The whole code is included in a single header file named \texttt{sharmonic.hpp}, that using codes can include directly.
However, the library has a build system based on CMake that will compile tests to ensure the correctness of the code.

The code of SHarmonic is designed to run on the GPU.
When compiled using Nvidia CUDA or AMD HIP, SHarmonic will include the necessary declarations so that its functions can be called from GPU kernels.
Additionally it can receive a complex type as template argument that can be used on the GPU (for example thrust::complex).

Since we are aware some researchers might need to implement the spherical harmonics on their own, in Appendix~\ref{sec:formulas}, we present all the formulas we use in SHarmonic. Additionally, Appendix~\ref{sec:reference_values} contains reference values that can be used to validate a spherical harmonics implementation.


\section{Conclusion}

In this article we have presented a new implementation of the spherical harmonics targeted for electronic-structure calculations.
This new library is based on the explicit implementation of the spherical-harmonics formulas in normalized Cartesian coordinates.
By writing the formulas in a form suited for finite precision calculations, our code results as accurate as existing implementations.
While at the same time it can be orders of magnitude faster than those same implementations.

The main limitation of our approach is that it is restricted to a maximum fixed value of \(\ell\).
This is probably not an issue for electronic structure, but it might prevent the library to be used for other scientific applications.
However, given the speed and accuracy of our implementation it could be easily extended using recurrence relationships for higher values of \(\ell\).

We have designed the library so it can be easily adopted by electronic structure codes. We expect it to be integrated into existing codes, especially when porting to \acp{GPU}, and directly used by new codes.
\acknowledgments{
We acknowledge support from the Computational Materials Sciences Program funded by the US Department of Energy, Office of Science, Basic Energy Sciences, Materials Sciences and Engineering Division for the materials application and the code development.
Part of this work was performed under the auspices of the U.S. Department of Energy by Lawrence Livermore National Laboratory under Contract No. DE-AC52-07NA27344.
}
\appendix

\section{Alternative implementation of the spherical harmonics}
\label{sec:alternatives}

In this appendix we discuss some alternative options for the implementation of the spherical harmonics that are currently available and that we use in this work for validation of our new library.

In C++, since version 17 the standard library provides the \texttt{std::sph\_legendre} function.
It calculates the associated Legendre polynomials and can be used to implement the spherical harmonics.
It has a few shortcomings, the most obvious one is that it is not available in other languages like C or Fortran.
We have also found that this function is not available in some C++ compilers since it is relatively new. A confusing aspect of this function is that it receives unsigned values of \(m\). This means that if the code passes a negative value, it might be silently converted into a very large positive integer and the function will just return zero. Instead, for negative values the function must be called with \(|m|\) and an additional \((-1)^m\) sign must be included.

Another alternative is the spherical harmonics implementation in the Boost library~\cite{Schling2014}.
The \texttt{boost::math::spherical\_harmonic} function has the advantage of directly calculating the complex harmonics.
However, it can only be called from C++ and introduces the additional dependency of Boost, a fairly large library that it is not always available in compilation environments. As out results show, this is also the slowest spherical-harmonics implementation tested.

Finally, we compare with an implementation based on the \ac{GSL}.
This is a C library, so it is more flexible to be used from other languages.
In fact, it is what the Octopus code~\cite{Andrade2015,TancogneDejean2020}, written in Fortran, uses.
Like the standard C++ function, GSL only provides the associated Legendre polynomials for positive \(m\) through the \texttt{gsl\_sf\_legendre\_Plm} function.
It is up to the user to implement the full spherical harmonics.
The biggest limitation of GSL has to do with its license.
It uses the \ac{GPL}, that legally limits its use to only other \ac{GPL} codes.

The major drawback of all these options for modern scientific codes is that none of them can be used when compiling code to run on \acp{GPU} using Nvidia CUDA or AMD HIP.

\section{Connection between \(m\) and \(-m\) for real harmonics}
\label{sec:mconnection}

To avoid implementing both the positive and negative harmonics for each value of \(m\) we derived formulas that connect both.
For complex harmonics this is trivial, since they are essentially complex conjugated, as show in eq.~(\ref{eq:conjm}).
For real harmonics, especially in Cartesian coordinates, it is not that simple.

From eq.~(\ref{eq:real_harmomics}), it is easy to see that the azimuthal part of the real harmonics \(\rY{\ell}{m}\) has the form \(\cos(m\phi)\), for \(m>0\), and \(\sin(m\phi)\) for \(m<0\).
Since \(\cos(x) = \sin(x + \pi/2)\), the spherical harmonics for  \(m\) and \(-m\) in terms of angular coordinates can be connected by a \(\pi/2m\) rotation
\begin{equation}
  \label{eq:rot}
  \rY{\ell}{m}(\theta, \phi) = \rY{\ell}{-m}\left(\theta, \phi + \frac{\pi}{2m}\right)\ .
\end{equation}
In Cartesian coordinates we can use this formula by rotating the \(x\) and \(x\) components using a 2-dimensional rotation operation, as
\begin{multline}
  \label{eq:rsymm}
  \rY{\ell}{m}(x, y, z) = \rY{\ell}{-m}\left(
  \cos\left(\frac{\pi}{2m}\right)x - \sin\left(\frac{\pi}{2m}\right)y,
  \right.\\
  \left.
  \sin\left(\frac{\pi}{2m}\right)x + \cos\left(\frac{\pi}{2m}\right)y,
  z\right)\ .
\end{multline}
In principle, we can use this relation to connect negative and positive \(m\) harmonics.
However, we would like to avoid the run-time calculation of the rotation coefficients that involve the slow evaluation of a division and trigonometric functions.
We use a different approach instead.

We can generalize eq.~(\ref{eq:rot}) by noting that we can connect \(\sin\) and \(\cos\) with any shift of the form \(\pi/2 + n\pi\), for any integer value \(n\).
We just need to take into account the sign for odd \(n\) that comes from \(\cos(x) = -\sin(x + 3\pi/2)\).
In general, we have that
\begin{equation}
  \label{eq:genrot}
  \rY{\ell}{m}(\theta, \phi) = (-1)^n\rY{\ell}{-m}\left(\theta, \phi + \frac{\pi}{2m}\left(1+ 2n\right)\right)\ .
\end{equation}
Now, if we pick \(n\) such that \(m=1 + 2n\), the shift simply becomes \(\pi/2\).
This tell us that for odd values of \(m\) we can use a simpler formula that always involves a \(\pi/2\) rotation:
\begin{equation}
  \label{eq:rsymm90}
  \rY{\ell}{2n + 1}(x, y, z) = (-1)^{n}\rY{\ell}{-(2n + 1)}(-y, x, z) \ .
  \end{equation}
We can do something similar for the case when \(m/2\) is odd.
In eq.~\ref{eq:genrot} we now pick \(n\) such that \(m = 2(1 + 2n)\).
This yields a formula for \(m=2,6,10,14,\dots\) through a \(\pi/4\) rotation
\begin{equation}
  \label{eq:rsymm45}
  \rY{\ell}{2(2n + 1)}(x, y, z) = (-1)^{n}\rY{\ell}{-2(2n + 1)}(\frac{x - y}{\sqrt{2}}, \frac{x + y}{\sqrt{2}}, z)\ .
\end{equation}
It is easy to see that we could continue deriving formulas for values of \(m\) that are higher powers of 2 times an odd number.
However for our implementation up \(\ell = 9)\) most values are covered by eq.~(\ref{eq:rsymm90}) and eq.~(\ref{eq:rsymm45}) already.
For the remaining values, \(m = 4\) and \(m = 8\), we use eq.~(\ref{eq:rsymm}) with hard-coded arguments for \(\cos\) and \(sin\) so that they are calculated at compile time rather than run time.

\section{Factorized formulas for spherical harmonics up to \(\ell=9\)}
\label{sec:formulas}

We now present the formulas for the spherical harmonics in \emph{normalized} cartesian coordinates, \(\xn\),\(\yn\) and \(\zn\), for \(l=0\) to \(l=9\) with \(m\) from \(-\ell\) to \(0\).
The values for positive \(m\) can be calculated using eqs.~(\ref{eq:rsymm}), (\ref{eq:rsymm90}) and (\ref{eq:rsymm45}).

The normalization condition, \(\xn^2 + \yn^2 = 1 - \zn^2\), has been used to simplify and factorize the formulas. This allows to write many polynomials in terms of \(\zn\) instead of \(\xn\) and \(\yn\). Note that these formulas will not work for any arbitrary set of values for \(x\), \(y\) and \(z\) that are not normalized.

For certain cases it is not possible to obtain simple analytical formulas for the root coefficients.
In those cases we obtain a high-precision numerical values for the roots, that we include in this table.
We also provide the polynomial these roots come from, however this form should be avoided for the numerical implementation as it is prone to large round-off errors.

The formulas use the \(\rpair\) function, defined in eq.~(\ref{eq:rpair}), that represents a polynomial factors of a pair of positive and negative roots.

\onecolumngrid

\begin{fleqn}

\subsection*{Spherical harmonics formula for \(\ell=0\)}
\begin{align*}
  \rY{0}{ 0}   =& \frac{1}{\sqrt{4\pi}}\\
   \end{align*}
   \subsection*{Spherical harmonics formulas for \(\ell=1\)}
 \begin{align*}
  \rY{1}{-1}	  =& \sqrt{\frac{3}{4\pi}}\yn\\
  \rY{1}{0} 	  =& \sqrt{\frac{3}{4\pi}}\zn\\
     \end{align*}
   \subsection*{Spherical harmonics formulas for \(\ell=2\)}
 \begin{align*}
	\rY{2}{-2}	  =& \frac{1}{2}\sqrt{\frac{15}{\pi}}\xn\yn\\
  \rY{2}{-1}   =& \frac{1}{2}\sqrt{\frac{15}{\pi}}\yn\zn\\
  \rY{2}{ 0}	  =& \frac{1}{4}\sqrt{\frac{5}{\pi}}\rpair\left(\sqrt{3}\zn, 1\right)\\
   \end{align*}
   
\subsection*{Spherical harmonics formulas for \(\ell=3\)}
 \begin{align*}
  \rY{3}{-3}	  =& \frac{1}{8}\sqrt{\frac{70}{\pi}}\yn\rpair\left(\sqrt{3}\xn, \yn\right)\\
  \rY{3}{-2}	  =& \frac{1}{2}\sqrt{\frac{105}{\pi}}\xn\yn\zn\\
  \rY{3}{-1}	  =& \frac{1}{8}\sqrt{\frac{42}{\pi}}\yn\rpair\left(\sqrt{5}\zn, 1\right)\\
  \rY{3}{ 0}	  =& \frac{1}{4}\sqrt{\frac{7}{\pi}}\rpair\left(\sqrt{5}\zn, \sqrt{3}\right)\zn\\
 \end{align*}
 
\subsection*{Spherical harmonics formulas for \(\ell=4\)}
 \begin{align*}
  \rY{4}{-4}	  =& \frac{3}{4}\sqrt{\frac{35}{\pi}}\xn\yn\rpair(\xn, \yn)\\
  \rY{4}{-3}	  =& \frac{3}{8}\sqrt{\frac{70}{\pi}}\yn\rpair\left(\sqrt{3}\xn, \yn\right)\zn\\
  \rY{4}{-2}	  =& \frac{3}{4}\sqrt{\frac{5}{\pi}}\xn\yn\rpair\left(\sqrt{7}\zn, 1\right)\\
  \rY{4}{-1}	  =& \frac{3}{8}\sqrt{\frac{10}{\pi}}\yn\zn\rpair\left(\sqrt{7}\zn, \sqrt{3}\right)\\
  \rY{4}{ 0}	  =& \frac{105}{16}\sqrt{\frac{1}{\pi}}\rpair\left(\sqrt{\frac{3 + 2\sqrt{\frac{6}{5}}}{7}}, \zn\right)\rpair\left(\sqrt{\frac{3 - 2\sqrt{\frac{6}{5}}}{7}}, \zn\right)\\
       \end{align*}
       
\subsection*{Spherical harmonics formulas for \(\ell=5\)}
 \begin{align*}
  \rY{5}{-5}	  =& \frac{30}{64}\sqrt{\frac{154}{\pi}}\yn\rpair\left(\sqrt{1 - \frac{2}{\sqrt{5}}}\yn, \xn\right)\rpair\left(\sqrt{1 + \frac{2}{\sqrt{5}}}\yn, \xn\right)\\
  \rY{5}{-4}	  =& \frac{3}{4}\sqrt{\frac{385}{\pi}}\rpair(\xn, \yn)\yn\zn\xn\\
  \rY{5}{-3}	  =& \frac{1}{32}\sqrt{\frac{770}{\pi}}\yn\rpair\left(\yn, \sqrt{3}\xn\right)\rpair\left(1, 3\zn\right)\\
  \rY{5}{-2}	  =& \frac{1}{4}\sqrt{\frac{1155}{\pi}}\xn\yn\rpair\left(\sqrt{3}\zn, 1\right)\zn\\
  \rY{5}{-1}	  =& \frac{21}{16}\sqrt{\frac{165}{\pi}}\yn\rpair\left(\sqrt{\frac{7 - 2\sqrt{7}}{21}}, \zn\right)\rpair\left(\sqrt{\frac{7 + 2\sqrt{7}}{21}}, \zn\right)\\
  \rY{5}{ 0}	  =& \frac{63}{16}\sqrt{\frac{11}{\pi}}\zn\rpair\left(\sqrt{\frac{5 - 2\sqrt{\frac{10}{7}}}{3}}, \zn\right)\rpair\left(\sqrt{\frac{5 + 2\sqrt{\frac{10}{7}}}{3}}, \zn\right)\\
     \end{align*}

\subsection*{Spherical harmonics formulas for \(\ell=6\)}
 \begin{align*}
  \rY{6}{-6}	  =& \frac{1}{32}\sqrt{\frac{6006}{\pi}}\xn\yn\rpair\left(\xn, \sqrt{3}\yn\right)\rpair\left(\sqrt{3}\xn, \yn\right)\\
  \rY{6}{-5}	  =& \frac{3}{32}\sqrt{\frac{2002}{\pi}}\yn\zn\rpair\left(\sqrt{5 - 2\sqrt{5}}\xn, \yn\right)\rpair\left(\sqrt{5 + 2\sqrt{5}}\xn, \yn\right)\\
  \rY{6}{-4}	  =& \frac{3}{8}\sqrt{\frac{91}{\pi}}\xn\yn\rpair(\xn, \yn)\rpair\left(\sqrt{11}\zn, 1\right)\\
  \rY{6}{-3}	  =& \frac{1}{32}\sqrt{\frac{2730}{\pi}}\yn\zn\rpair\left(\sqrt{3}\xn, \yn\right)\rpair\left(\sqrt{11}\zn, \sqrt{3}\right)\\
  \rY{6}{-2}	  =& \frac{33}{32}\sqrt{\frac{2730}{\pi}}\xn\yn\rpair\left(\sqrt{\frac{3 - \frac{4}{\sqrt{3}}}{11}}, \zn\right)\rpair\left(\sqrt{\frac{3 + \frac{4}{\sqrt{3}}}{11}}, \zn\right)\\
  \rY{6}{-1}	  =& \frac{33}{16}\sqrt{\frac{273}{\pi}}\yn\zn\rpair\left(\sqrt{\frac{15 - 2\sqrt{15}}{33}}, \zn\right)\rpair\left(\sqrt{\frac{15 + 2\sqrt{15}}{33}}, \zn\right)\\
  \rY{6}{ 0}	  =& \frac{1}{32}\sqrt{\frac{13}{\pi}}\left(\zn^2\left(\left(231\zn^2 - 315\right)\zn^2 + 105\right) - 5\right)\\
\end{align*}
       
\subsection*{Spherical harmonics formulas for \(\ell=7\)}
 \begin{align*}
  \rY{7}{-7}	  =& \frac{3}{64}\sqrt{\frac{715}{\pi}}\yn\left(7\xn^6 - 35\xn^4\yn^2 + 21\xn^2\yn^4 - \yn^6\right)\\  
                        =& \frac{3}{64}\sqrt{\frac{715}{\pi}}\yn\\
                        & \times\rpair(0.481574618807528644332162353056970575219\,\xn, \yn)\\
                        & \times\rpair(1.253960337662703837570910978336464443221\,\xn, \yn)\\
                        & \times\rpair(4.381286267534823072404689085032695444150\,\xn, \yn)\\
  \rY{7}{-6}	  =& \frac{3}{32}\sqrt{\frac{10010}{\pi}}\xn\yn\zn\rpair\left(\xn, \sqrt{3}\yn\right)\rpair\left(\sqrt{3}\xn, \yn\right)\\
  \rY{7}{-5}	  =& \frac{3}{64}\sqrt{\frac{385}{\pi}}\yn\rpair\left(\sqrt{13}\zn, 1\right)\rpair\left(\sqrt{5 - 2\sqrt{5}}\xn,\yn\right)\rpair\left(\sqrt{5 + 2\sqrt{5}}\xn, \yn\right)\\
  \rY{7}{-4}	  =& \frac{3}{8}\sqrt{\frac{385}{\pi}}\xn\yn\zn\rpair(\xn, \yn)\rpair\left(\sqrt{13}\zn, \sqrt{3}\right)\\
  \rY{7}{-3}	  =& \frac{429}{64}\sqrt{\frac{35}{\pi}}\yn\rpair\left(\sqrt{3}\xn, \yn\right)\rpair\left(\sqrt{\frac{33 - 2\sqrt{165}}{143}}, \zn\right)\rpair\left(\sqrt{\frac{33 + 2\sqrt{165}}{143}}, \zn\right)\\
  \rY{7}{-2}	  =& \frac{429}{32}\sqrt{\frac{70}{\pi}}\xn\yn\zn\rpair\left(\sqrt{\frac{55 - 4\sqrt{55}}{143}}, \zn\right)\rpair\left(\sqrt{\frac{55 + 4\sqrt{55}}{143}}, \zn\right)\\
  \rY{7}{-1}	  =& \frac{1}{64}\sqrt{\frac{105}{\pi}}\yn\left(\zn^2\left(\left(429\zn^2 - 495\right)\zn^2 + 135\right) - 5\right)\\
  \rY{7}{ 0}	  =& \frac{1}{32}\sqrt{\frac{15}{\pi}}\zn\left(-35 + \zn^2\left(315 + \zn^2\left(-693 + 429\zn^2\right)\right)\right)\\
\end{align*}

\subsection*{Spherical harmonics formulas for \(\ell=8\)}
 \begin{align*}
  \rY{8}{-8}	  =& \frac{12}{128}\sqrt{\frac{12155}{\pi}}\xn\yn\rpair(\xn, \yn)\rpair\left(\left(1 + \sqrt{2}\right)\xn, \yn\right)\rpair\left(\left(1 - \sqrt{2}\right)\xn, \yn\right)\\
  \rY{8}{-7}	  =& \frac{3}{64}\sqrt{\frac{12155}{\pi}}\yn\zn\left(7\xn^6 - 35\xn^4\yn^2 + 21\xn^2\yn^4 - \yn^6\right)\\
                        =& \frac{3}{64}\sqrt{\frac{12155}{\pi}}\yn\zn\\
                        &\times\rpair(0.481574618807528644332162353056970575219\, \xn, \yn)\\
                        &\times\rpair(1.253960337662703837570910978336464443221\, \xn, \yn)\\
                        &\times\rpair(4.381286267534823072404689085032695444150\, \xn, \yn)\\
  \rY{8}{-6}	  =& \frac{1}{64}\sqrt{\frac{14586}{\pi}}\xn\yn\rpair\left(\xn, \sqrt{3}\yn\right)\rpair\left(\sqrt{3}\xn, \yn\right)\rpair\left(\sqrt{15}\zn, 1\right)\\
  \rY{8}{-5}	  =& \frac{3}{64}\sqrt{\frac{17017}{\pi}}\yn\zn\rpair\left(\sqrt{5}\zn, 1\right)\rpair\left(\sqrt{5 + 2\sqrt{5}}\xn, \yn\right)\rpair\left(\sqrt{5 - 2\sqrt{5}}\xn, \yn\right)\\
  \rY{8}{-4}	  =& \frac{195}{32}\sqrt{\frac{1309}{\pi}}\xn\yn\rpair(\xn, \yn)\rpair\left(\sqrt{\frac{13 - 2\sqrt{26}}{65}}, \zn\right)\rpair\left(\sqrt{\frac{13 + 2\sqrt{26}}{65}}, \zn\right)\\
  \rY{8}{-3}	  =& \frac{39}{64}\sqrt{\frac{19635}{\pi}}\yn\zn\rpair\left(\sqrt{3}\xn, \yn\right)\rpair\left(\sqrt{\frac{1 - \frac{2}{\sqrt{13}}}{3}}, \zn\right)\rpair\left(\sqrt{\frac{1 + \frac{2}{\sqrt{13}}}{3}}, \zn\right)\\
  \rY{8}{-2}	  =& \frac{3}{64}\sqrt{\frac{1190}{\pi}}\xn\yn\left(\zn^2\left(\left(143\zn^2 - 143\right)\zn^2 + 33\right) - 1\right)\\
  \rY{8}{-1}	  =& \frac{3}{64}\sqrt{\frac{17}{\pi}}\yn\zn\left(\zn^2\left(\left(715\zn^2 - 1001\right)\zn^2 + 385\right) - 35\right)\\
                        =& \frac{2145}{64}\sqrt{\frac{17}{\pi}}\yn\zn\\
                        & \times\rpair(0.36311746382617815871075206870865921, \zn)\\
                        & \times\rpair(0.67718627951073775344588542709134245, \zn)\\
                        & \times\rpair(0.89975799541146015731234524441833796, \zn)\\
  \rY{8}{ 0}	  =& \frac{1}{256}\sqrt{\frac{17}{\pi}}\left(\left(\zn^2\left(\left(6435\zn^2 - 12012\right)\zn^2 + 6930\right) - 1260\right)\zn^2 + 35\right)\\
\end{align*}

\subsection*{Spherical harmonics formulas for \(\ell=9\)}
\begin{align*}
  \rY{9}{-9}	  =& \frac{1}{512}\sqrt{\frac{461890}{\pi}}\yn\rpair\left(\sqrt{3}\xn, \yn\right)\left(3\xn^6 - 27\xn^4\yn^2 + 33\xn^2\yn^4 - \yn^6\right)\\
                        =& \frac{1}{512}\sqrt{\frac{461890}{\pi}}\yn\rpair\left(\yn, \sqrt{3}\xn\right)\\
                        &  \times\rpair(\yn, 0.363970234266202361351047882776834043890\,\xn)\\
                        &  \times\rpair(\yn, 0.839099631177280011763127298123181364687\,\xn)\\
                        &  \times\rpair(\yn, 5.671281819617709530994418439863964421625\,\xn)\\
	\rY{9}{-8}	  =& \frac{3}{32}\sqrt{\frac{230945}{\pi}}\xn\yn\zn\rpair(\xn, \yn)\rpair\left(\left(1 + \sqrt{2}\right)\xn, \yn\right)\rpair\left(\left(1 - \sqrt{2}\right)\xn, \yn\right)\\
  \rY{9}{-7}	  =& -\frac{3}{512}\sqrt{\frac{27170}{\pi}}\rpair\left(1,\sqrt{17}\zn\right)\left(7\xn^6 - 35\xn^4\yn^2 + 21\xn^2\yn^4 - \yn^6\right)\\
                        =& \frac{3}{512}\sqrt{\frac{27170}{\pi}}\yn\rpair\left(1, \sqrt{17}\zn\right)\\
                        & \times\rpair(\yn, 0.481574618807528644332162353056970575219\,\xn)\\
                        & \times\rpair(\yn, 1.253960337662703837570910978336464443221\,\xn)\\
                        & \times\rpair(\yn, 4.381286267534823072404689085032695444150\,\xn)\\
  \rY{9}{-6}	  =& \frac{1}{64}\sqrt{\frac{81510}{\pi}}\xn\yn\zn\rpair\left(\xn, \sqrt{3}\yn\right)\rpair\left(\sqrt{3}\xn, \yn\right)\rpair\left(\sqrt{17}\zn, \sqrt{3}\right)\\
  \rY{9}{-5}	  =& \frac{255}{256}\sqrt{\frac{5434}{\pi}}\yn\rpair\left(\sqrt{5 - 2\sqrt{5}}\xn, \yn\right)\rpair\left(\sqrt{5 + 2\sqrt{5}}\xn, \yn\right)\rpair\left(\sqrt{\frac{15 - 2\sqrt{35}}{85}}, \zn\right)\rpair\left(\sqrt{\frac{15 + 2\sqrt{35}}{85}}, \zn\right)\\
  \rY{9}{-4}	  =& \frac{51}{32}\sqrt{\frac{95095}{\pi}}\xn\yn\zn\rpair(\xn, \yn)\rpair\left(\sqrt{\frac{5 - 2\sqrt{2}}{17}}, \zn\right)\rpair\left(\sqrt{\frac{5 + 2\sqrt{2})}{17}}, \zn\right)\\
  \rY{9}{-3}	  =& \frac{1}{256}\sqrt{\frac{43890}{\pi}}\yn\rpair\left(\sqrt{3}\xn, \yn\right)\left(\zn^2\left(\left(221\zn^2 - 195\right)\zn^2 + 39\right) - 1\right)\\
  \rY{9}{-2}	  =& -\frac{3}{64}\sqrt{\frac{2090}{\pi}}\xn\yn\zn\left(\left(\zn^2\left(273 - 221\zn^2\right) - 91\right)\zn^2 + 7\right)\\
  \rY{9}{-1}	  =& \frac{3}{256}\sqrt{\frac{95}{\pi}}\yn\left(\left(\zn^2\left(\left(2431\zn^2 - 4004\right)\zn^2 + 2002\right) - 308\right)\zn^2 + 7\right)\\
                        =& \frac{7293}{256}\sqrt{\frac{95}{\pi}}\yn\\
                        &  \times\rpair(0.16527895766638702462621976595817353, \zn)
                        \rpair(0.47792494981044449566117509273125800, \zn)\\
                        &  \times\rpair(0.73877386510550507500310617485983073, \zn)
                        \rpair(0.91953390816645881382893266082233813, \zn)\\
  \rY{9}{0}	  =& \frac{1}{256}\sqrt{\frac{19}{\pi}}\zn\left(\left(\zn^2\left(\left(12155\zn^2 - 25740\right)\zn^2 + 18018\right) - 4620\right)\zn^2 + 315\right)\\
                        =& \frac{12155}{256}\sqrt{\frac{19}{\pi}}\zn\\
                        &  \times\rpair(0.32425342340380892903853801464333661, \zn)
                        \rpair(0.61337143270059039730870203934147418, \zn)\\
                        & \times\rpair(0.83603110732663579429942978806973488, \zn)
                         \rpair(0.96816023950762608983557620290367287, \zn)\\
\end{align*}
\end{fleqn}

\twocolumngrid

\section{Reference values for the spherical harmonics}
\label{sec:reference_values}

In tables~\ref{tab:ref1} and~\ref{tab:ref2} we provide values calculated with SHarmonic for two points defined by their angles \(\theta\) and \(\phi\).
The purpose of these tables is to provide a reference that can be useful to validate an implementation of the spherical harmonics.
These values match calculations obtained with other implementations as well.
Cartesian coordinates can be obtained from \(\theta\) and \(\phi\) as
\begin{align*}
	x &= \sin\theta\cos\phi\\
	y &= \sin\theta\sin\phi\\
	z &= \cos\theta\ .
\end{align*}
(This is the inverse of eqs.~(\ref{eq:spherical_theta})) and (\ref{eq:spherical_phi})).

\begin{table}
      \centering
      \scalebox{0.85}{   
    \begin{tabular}{ll|ll}
    \(\ell\) & \(m\) & \( \rY{l}{-m}(\theta, \phi)\) & \(\rY{l}{m}(\theta, \phi)\)\\
    \hline
0 &  0 & 2.82094791773878140e-01 & \\
1 &  0 & -2.55722001017027356e-01 & \\
1 &  1 & -1.05634976792300384e-01 & -4.02715686945245066e-01\\
2 &  0 & -5.62147632675229145e-02 & \\
2 &  1 & 1.23624669105401527e-01 & 4.71298381028178615e-01\\
2 &  2 & -3.45570896358408208e-01 & 1.94686391253623081e-01\\
3 &  0 & 3.18434249038601458e-01 & \\
3 &  1 & -3.65214185609349096e-02 & -1.39231801914437098e-01\\
3 &  2 & 4.78518329848811652e-01 & -2.69585800681415744e-01\\
3 &  3 & 2.54019229733874252e-01 & 2.62184202761883589e-01\\
4 &  0 & -2.74141274717346561e-01 & \\
4 &  1 & -8.19541375512665149e-02 & -3.12436446754339070e-01\\
4 &  2 & -2.74566575097891730e-01 & 1.54684252140363465e-01\\
4 &  3 & -3.98841414904085589e-01 & -4.11661426202272196e-01\\
4 &  4 & 1.70955953966685115e-01 & -2.82190055112113369e-01\\
5 &  0 & -3.38241558580449284e-02 & \\
5 &  1 & 1.23309330674236661e-01 & 4.70096205983671467e-01\\
5 &  2 & -1.41437716730428292e-01 & 7.96826322690101785e-02\\
5 &  3 & 3.08620906166296116e-01 & 3.18540947957492715e-01\\
5 &  4 & -2.96751533654975974e-01 & 4.89835713197919187e-01\\
5 &  5 & -2.82703168796350934e-01 & -8.37958855096906235e-02\\
6 &  0 & 3.09846245017294242e-01 & \\
6 &  1 & -4.88865289200799136e-02 & -1.86371717723078645e-01\\
6 &  2 & 4.23805057928605844e-01 & -2.38761649758777172e-01\\
6 &  3 & -2.72475373952341940e-03 & -2.81233585216301811e-03\\
6 &  4 & 2.77467519049719724e-01 & -4.58004372914207813e-01\\
6 &  5 & 5.33475834457012299e-01 & 1.58127268741543286e-01\\
6 &  6 & -8.27075588763272716e-03 & 2.61380191337153900e-01\\
7 &  0 & -2.89647176025430497e-01 & \\
7 &  1 & -7.13883837177109321e-02 & -2.72156276848675382e-01\\
7 &  2 & -3.22453348860354594e-01 & 1.81662517008257168e-01\\
7 &  3 & -2.92943223552905863e-01 & -3.02359335559675813e-01\\
7 &  4 & -8.32341531063935797e-02 & 1.37391238546043654e-01\\
7 &  5 & -5.72367474797551234e-01 & -1.69655117739965722e-01\\
7 &  6 & 1.67649873881012233e-02 & -5.29822868768178079e-01\\
7 &  7 & 2.24846914334883241e-01 & -5.14371474445672397e-02\\
8 &  0 & -7.43735782960527891e-03 & \\
8 &  1 & 1.23966082516397832e-01 & 4.72599962573579391e-01\\
8 &  2 & -7.17947446387538907e-02 & 4.04474447703423029e-02\\
8 &  3 & 3.32786929681476051e-01 & 3.43483743098987715e-01\\
8 &  4 & -1.62681278350457953e-01 & 2.68531383893016617e-01\\
8 &  5 & 2.87428672353492343e-01 & 8.51965693320042572e-02\\
8 &  6 & -2.00347913914888306e-02 & 6.33158284255222026e-01\\
8 &  7 & -4.85203350770030906e-01 & 1.10997637516985950e-01\\
8 &  8 & -9.38078598619499354e-02 & -1.79563908384784060e-01\\
9 &  0 & 2.97539368635582557e-01 & \\
9 &  1 & -5.90690698905085368e-02 & -2.25190952660925120e-01\\
9 &  2 & 3.97311925510614172e-01 & -2.23836051573814004e-01\\
9 &  3 & -8.07468972106639843e-02 & -8.33423551943419660e-02\\
9 &  4 & 2.70299426419996103e-01 & -4.46172293321211333e-01\\
9 &  5 & 1.76296777532887639e-01 & 5.22560275810531882e-02\\
9 &  6 & 1.32090473198986386e-02 & -4.17444712764316717e-01\\
9 &  7 & 6.33537449515871320e-01 & -1.44931316041401775e-01\\
9 &  8 & 2.14007220903021045e-01 & 4.09645557040358099e-01\\
9 &  9 & -1.31218772176712128e-01 & 1.19322150190607823e-01\\
    \end{tabular}
    }
    \caption{Reference values for the real spherical harmonics up to \(\ell = 9\) for the point \(\theta=2.12160245947564796\) and \(\phi=-1.82732370250979703\).}
    \label{tab:ref1}
\end{table}

\begin{table}
      \centering
      \scalebox{0.85}{   
    \begin{tabular}{ll|ll}
    \(\ell\) & \(m\) & \( \rY{l}{-m}(\theta, \phi)\) & \(\rY{l}{m}(\theta, \phi)\)\\
    \hline
0 &  0 & 2.82094791773878140e-01 & \\
1 &  0 & -5.83481414444863219e-02 & \\
1 &  1 & 3.52213965191625900e-01 & 3.33576425653566488e-01\\
2 &  0 & -3.01898395233452077e-01 & \\
2 &  1 & -9.40508891152187221e-02 & -8.90741495826994495e-02\\
2 &  2 & 2.92468625790618109e-02 & 5.37689095897659008e-01\\
3 &  0 & 1.30514795718817689e-01 & \\
3 &  1 & -3.05973819059547569e-01 & -2.89783094914834061e-01\\
3 &  2 & -9.24058861679355643e-03 & -1.69883649074994564e-01\\
3 &  3 & -3.73728868391523250e-01 & 4.40221375107542001e-01\\
4 &  0 & 2.72852146179700994e-01 & \\
4 &  1 & 1.67033159586032515e-01 & 1.58194534705676154e-01\\
4 &  2 & -2.28000978080050223e-02 & -4.19168515721103252e-01\\
4 &  3 & 1.33890336276837080e-01 & -1.57711627156542439e-01\\
4 &  4 & -6.04525430908636752e-01 & 6.59598202754702601e-02\\
5 &  0 & -1.95727655537335454e-01 & \\
5 &  1 & 2.62717387357419729e-01 & 2.48815594191593242e-01\\
5 &  2 & 1.46681962937466846e-02 & 2.69667530399667454e-01\\
5 &  3 & 2.70107539742010683e-01 & -3.18164109408787588e-01\\
5 &  4 & 2.39432031065408502e-01 & -2.61244489144609147e-02\\
5 &  5 & -5.04277710198387297e-01 & -3.82994564609737353e-01\\
6 &  0 & -2.26688512655598251e-01 & \\
6 &  1 & -2.29651174369361916e-01 & -2.17499092779007891e-01\\
6 &  2 & 1.84955137257718678e-02 & 3.40030867464486575e-01\\
6 &  3 & -1.98105831006451405e-01 & 2.33352113573200293e-01\\
6 &  4 & 4.10928859020655235e-01 & -4.48364821414814554e-02\\
6 &  5 & 2.17126474138350928e-01 & 1.64905681425320055e-01\\
6 &  6 & -1.06203816463220524e-01 & -6.45694857930299793e-01\\
7 &  0 & 2.49912413462847705e-01 & \\
7 &  1 & -2.06693957623301949e-01 & -1.95756666123839712e-01\\
7 &  2 & -1.90420274855719988e-02 & -3.50078252499657350e-01\\
7 &  3 & -2.06907729748927099e-01 & 2.43720014732793644e-01\\
7 &  4 & -3.36953909231299309e-01 & 3.67650691892440762e-02\\
7 &  5 & 3.24757794691705237e-01 & 2.46650739594728025e-01\\
7 &  6 & 4.91198559794929698e-02 & 2.98637464118054752e-01\\
7 &  7 & 3.77052895953012390e-01 & -5.56844104803174811e-01\\
8 &  0 & 1.67253102981562901e-01 & \\
8 &  1 & 2.78889291525095395e-01 & 2.64131755733712392e-01\\
8 &  2 & -1.36289019515063430e-02 & -2.50560618205577068e-01\\
8 &  3 & 2.46399332956732664e-01 & -2.90237823068457967e-01\\
8 &  4 & -2.96887870807171550e-01 & 3.23934603892084144e-02\\
8 &  5 & -2.93944891081749038e-01 & -2.23248605485301871e-01\\
8 &  6 & 6.50514630685824352e-02 & 3.95497983057630453e-01\\
8 &  7 & -1.85651333551225262e-01 & 2.74175988956548267e-01\\
8 &  8 & 6.72033568021081384e-01 & -1.48418197689477727e-01\\
9 &  0 & -2.89899747508637173e-01 & \\
9 &  1 & 1.39514888233566392e-01 & 1.32132403430085199e-01\\
9 &  2 & 2.22513789835169229e-02 & 4.09080591662805970e-01\\
9 &  3 & 1.41204801700664406e-01 & -1.66327456168928328e-01\\
9 &  4 & 4.04918912353927518e-01 & -4.41807363585345197e-02\\
9 &  5 & -2.20717226600193051e-01 & -1.67632826900783377e-01\\
9 &  6 & -6.44199418831517406e-02 & -3.91658479020147332e-01\\
9 &  7 & -2.20102608054930010e-01 & 3.25054763039043570e-01\\
9 &  8 & -3.49815540290367211e-01 & 7.72565456314207866e-02\\
9 &  9 & 6.01820889408594351e-01 & 3.61459549167042316e-01\\
    \end{tabular}
    }
    \caption{Reference values for the real spherical harmonics up to \(\ell = 9\) for the point \(\theta=1.69050041976591414\) and \(\phi=7.58228122208986166\).}
    \label{tab:ref2}
\end{table}

\bibliography{\jobname}

\newpage

\end{document}